\documentclass[twocolumn,showpacs,
			amsmath,amssymb,prl,,reprint,balancelastpage,
			superscriptaddress,raggedbottom]{revtex4-1}
\usepackage{graphicx}
\usepackage{bm}
\usepackage{dcolumn}
 \usepackage{xfrac}
\usepackage[usenames,dvipsnames,table]{xcolor}
\usepackage[breaklinks=true,colorlinks=true,linkcolor=RubineRed,urlcolor=blue,citecolor=blue]
{hyperref}
\usepackage{microtype}



\begin{document}

\title{Electronic correlations and screening effects in the Hund's polar metal SrEuMo$_2$O$_6$}
\author{Gianluca Giovannetti}
\email{ggiovann@sissa.it}
\affiliation{CNR-IOM-Democritos National Simulation Centre and International School
for Advanced Studies (SISSA), Via Bonomea 265, I-34136, Trieste, Italy}
\author{Danilo Puggioni}
\affiliation{Department of Materials Science and Engineering
Northwestern University
2220 Campus Drive, Evanston, IL 60208-3108, USA}
\author{James M.\ Rondinelli}
\email{jrondinelli@northwestern.edu}
\affiliation{Department of Materials Science and Engineering
Northwestern University
2220 Campus Drive, Evanston, IL 60208-3108, USA}
\author{Massimo Capone}
\affiliation{CNR-IOM-Democritos National Simulation Centre and International School
for Advanced Studies (SISSA), Via Bonomea 265, I-34136, Trieste, Italy}


\begin{abstract}
Using a first-principles approach based on density functional theory and dynamical mean field theory, we study the electronic properties of a new candidate polar metal SrEuMo$_2$O$_6$. Its electronic structure shares similarities with centrosymmetric SrMoO$_3$ and EuMoO$_3$, 
from which it may be considered an ordered derivative, but ferroelectric-like distortions of the divalent cations and oxygen anions lift inversion symmetry mediated by an anharmonic lattice interaction in the metallic state. 
We find that Hund's coupling promotes the effects of electronic correlations owing to the Mo$^{4+}$ $d^{2}$ electronic configuration, producing a correlated metallic phase far from the Mott state. The contraindication between metallicity and polar distortions is thereby  alleviated through the renormalized quasiparticles, which are unable to fully screen the ordered local dipoles. 
\end{abstract}

\pacs{75.85.+t, 71.20.-b, 71.45.Gm, 77.80.B-, 78.20.Ci}

\maketitle

``Ferroelectric metals'' are emerging as a new paradigm in condensed matter physics from which exotic phenomena can be discovered and new technology platforms established \cite{Keppen}.
The possible existence of these materials was predicted theoretically a half century ago \cite{Anderson}, but a conclusive observation of an intrinsic metallic material which undergoes an inversion-symmetry lifting (polar) transition has been obtained only recently in LiOsO$_3$ \cite{Shi}.
The theoretical appeal of these materials lies in the contrast with one's intuition: Itinerant electrons in metals are expected to screen electric fields and to inhibit long-range Coulomb interactions from cooperatively ordering ferroelectric-like displacements \cite{Cochran,Gonze,Ghosez}. It is therefore uncommon and surprising to find  metals that undergo the same structural transitions that occur in isostructural and insulating analogues; it is this  incompatibility, in part, that explains the scarcity of polar non-centrosymmetric metals (NCSM). 

Nonetheless, polar displacements and metallicity can coexist provided that the ferroelectric-like distortions are largely decoupled from the electronic structure at the Fermi level, \emph{i.e.}, the electrons responsible for transport \cite{Anderson,PuggioniRondinelli}. This appears to be the main operational principle active in oxide-based NCSM: LiOsO$_3$ \cite{Shi} exhibits a second-order displacive transition from a non-polar to polar structure (crystal class $C_{3v}$) with displacements of Li and O rather than Os, whose orbitals are responsible for the metallic behavior \cite{GGMC};  LaSr$_2$Cu$_2$GaO$_7$ has a polar structure ($C_{2v}$) owing to the presence of intrinsically acentric tetrahedral GaO$_4$ units, whereas the conduction bandi is mainly of copper character \cite{Poeppelmeier,Vaughey},  and  Cd$_2$Re$_2$O$_7$ is a geometrically frustrated pyrochlore that becomes optically active ($D_{2d}$)  as oxygen rather than Re displacements lift inversion symmetry below 200\,K \cite{Sergienko,Petersen}.

Although the framework described in Ref.~\onlinecite{PuggioniRondinelli} provides a  \emph{structural} understanding for the stability of oxide and non-oxide NCSM, it is partially limited, as it considers all metallic states equivalent assuming a Fermi liquid description.
Yet, the rich phenomenology of {\it{strongly correlated}} oxides demonstrates that strong electron-electron interactions can lead to metallic states which challenge the Landau Fermi-liquid paradigm.
Anomalous metallic states range from non-Fermi-liquid metals, with unconventional scaling behavior of the resistivity and other observables, to `bad' metals that retain Fermi-liquid coherence only below a very low coherence temperature, above which the Mott-Ioffe-Regel limit breaks \cite{resilient}. More recently the possibility of a `bad metal' far from the Mott-Hubbard localization state has been proposed in the presence of a sizable Hund's coupling \cite{GeorgesAnnuRev,Haule}.

Therefore, we pose the question: How does the nature of the metallicity dictate the compatibility between inversion lifting displacements and `delocalized' electrons? The crucial and non-trivial role of the quasiparticle character on screening of the long-range Coulomb interactions may be gleaned from recent studies on electron doped BaTiO$_3$  \cite{Wang3,Kolodiazhnyi}: Optical conductivity measurements show spectral weight shifts with carrier concentration \cite{Couillard} that are accompanied by a decrease in the ferroelectric-like distortion across the doping-driven insulator-metal transition.

As a starting point, we first discuss the previously mentioned NCSM oxides within this context. Indeed all may be considered to exhibit strong electron correlations. 
The residual resistivity in LiOsO$_3$ is  larger than that expected for a normal metal and the magnetic susceptibility displays a  Curie-Weiss component suggesting an incipient localization of the carriers \cite{Shi}. 
LaSr$_2$Cu$_2$GaO$_7$ shares the same Cu-O planar structure of YBa$_2$Cu$_3$O$_7$, one of the most studied cuprate superconductors for which strong deviations from the Fermi-liquid paradigm are well established \cite{Cuprate}. Optical studies of Cd$_2$Re$_2$O$_7$ reveal an exotic metallic state with strong mass renormalizations \cite{Wang}. Lastly, a recent theoretical proposal identifies 
NCSM in  ultra-short superlattices of  SrRuO$_3$ and CaRuO$_3$ \cite{PuggioniRondinelli}, two compounds displaying strong correlations due to the Hund's coupling \cite{MravljeIncoherent}.

Although the role of strong correlations in compounds with half-filled shells follows the longstanding Mott-Hubbard paradigm, materials with  integer $d$-orbital occupation that are  different from half-filling (like $d^{2}$ and $d^{4}$ configurations which \emph{partially} fill the three $t_{2g}$ bands) and a sizable Hund's coupling display unexpected properties. These features have  led to a class of materials, dubbed `Hund's metals', where sizable electron-electron correlations exist even for moderate values of the Coulomb repulsion very far from the critical value of the Mott transition \cite{MediciJanus,GeorgesAnnuRev}.
However, the extent to which the Hund's interaction  can drive a correlated metallic state compatible with ferroelectric-like displacements is unknown. 

In this Letter, we demonstrate the possibility to engineer polar distortions in a `Hund's metal', proposing SrEuMo$_2$O$_6$ (SEMO) as a new compound. To characterize this material and its electronic properties, we use a combination of density funcitonal theory (DFT) and dynamical mean-field theory (DMFT) \cite{DMFT}. Utilizing DFT calculations, we first show that the ordered molybdate exhibits a 
polar metallic ground state, fulfilling the structural criterion described in Ref.~\onlinecite{PuggioniRondinelli}. 
We then establish with our DFT+DMFT approach that SEMO is 
in proximity to region of phase space exhibiting bad metallic behavior 
driven and stabilized by the Hund's coupling 
owing to the $d^{2}$ electronic configuration in the Mo $4d$ manifold.
The enhanced electron correlations that characterize the poor-metallic state 
reduce the ability of the metal to screen the ferroelectric-like distortions, 
leading to a `ferroelectric' metallic state. 
Our finds indicate that electronic configurations conducive to 
Hund's-metal behavior may be a promising arena for the discovery of 
yet unknown noncentrosymmetric metals.

\begin{figure}
\includegraphics[width=.95\columnwidth,angle=-0]{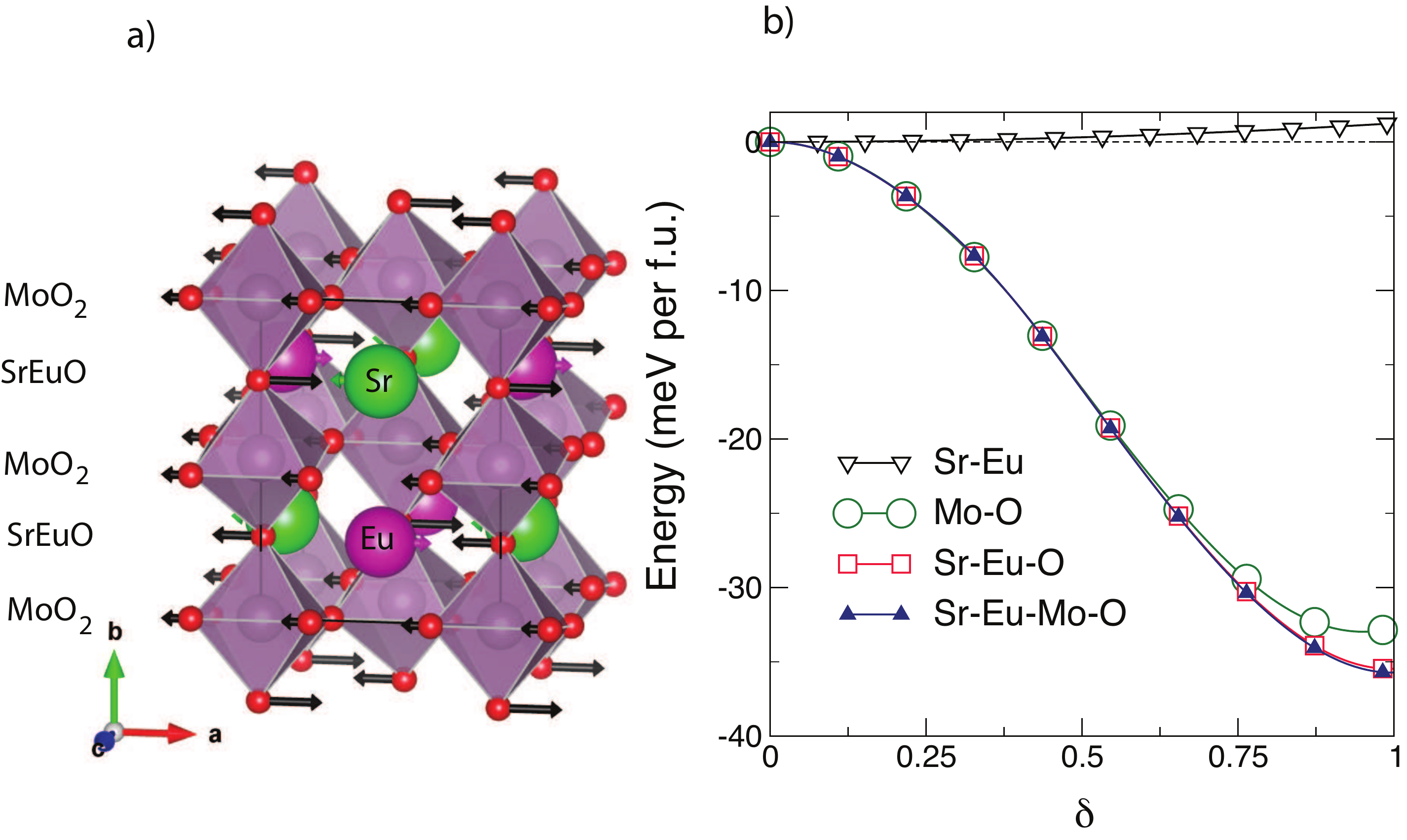}\vspace{-8pt}
\caption{(Color online) a) DFT relaxed crystal structure of SEMO in $Imm2$ ferroelectric crystal symmetry with polar displacements along the 
[100]-direction. b) Total energy gain (per formula unit, f.u.) as a function of ionic displacement given as a percentage present in the equilibrium structure ($\delta$). 
Energy of the full mode is compared to that of the partial modes, where a reduced set of atoms are displaced (see legend).}
\label{fig1}
\end{figure}

A promising route to lift inversion symmetry and realize a NCSM is to 
use cation ordering in combination with non-polar cooperative displacements \cite{PuggioniRondinelli}; the 
former being experimentally accessible through either synthetic bulk chemistry routes or heteroepitaxial thin film 
growth \cite{King,Zubko}.
With this in mind, we select SrMoO$_3$ (SMO) and EuMoO$_3$ (EMO) with $d^{2}$ electronic configurations and nearly identical pseudo-cubic lattice constants ($a$=3.975\,\AA)  \cite{Kozuka,Macquart}. We then construct a 1/1 period superlattice of SMO and EMO in which the Sr and Eu cations are arranged in  a rock-salt structure (see Fig. \ref{fig1}a) 
and the in-plane lattice parameters are fixed to the pseudo-cubic lattice constant,  $a$=$c$=3.975\,\AA.
This cation ordering should be experimentally accessible via growth on a (111) terminated perovskite substrate.

Next variable-volume atomic relaxations along the 
$b$ direction are performed starting from a centrosymmetric $I4/mmm$ symmetry with  spin-polarized DFT calculations within the generalized-gradient approximation (PBE) \cite{PBE} as implemented in the Vienna {\it Ab initio} Simulation Package (VASP) \cite{VASP} with the projector augmented wave (PAW) method \cite{PAW,note1}.
Spin-polarized calculations for SEMO lead to a 
non-magnetic ground state, consistent with that of bulk SMO and EMO.

We determine the most stable ionic configuration to be polar with $Imm2$ symmetry and 35 meV per formula unit (f.u.) lower in energy than the centric $I4/mmm$ phase  \cite{relaxed_structure}.
The octahedral tilt pattern of SMO and EMO is $a^{0}a^{0}a^{0}$ and $a^{0}a^{0}c^{-}$, respectively, whereas in SEMO we find the $a^{-}b^{+}a^{-}$ tilt  with $a^{-}$ and $b^{+}$ rotations of 4$^{\circ}$ and 0.11$^{\circ}$, respectively. 
%
The loss of inversion symmetry is a consequence of the out-of-phase octahedral tilt modes (see \autoref{fig1}a), which allow for anti-polar Sr and Eu cation displacements connected with neighboring O sites \cite{Young}.
(Note that if the $b^+$ tilt amplitude is zero, the structure remains polar 
owing to the out-of-phase rotations that in combination with the rock salt 
Sr and Eu order lift inversion.)
Along the polar  $a$ axis (see \autoref{fig1}a) we find ($i$) Sr and Eu ions have opposite displacements of respectively 0.045, -0.018\,\AA\; ($ii$) apical O ions belonging to the same octahedra displace 0.33 and -0.28\,\AA\, while planar O ions displace by -0.020\,\AA\; and ($iv$) the Mo cations have negligible ferroelectric displacements.
The situation along the non-polar $b$ and $c$ axes is different: the Sr, Eu and apical O ions do not displace while the planar O sites exhibit antipolar displacements.

In \autoref{fig1}b we show the variation of the total energy as a function of the 
amplitude  $\delta$ of atomic displacements for the different atoms involved in the 
distortion mode connecting the centrosymmetric ($\delta = 0$) structure to the polar ground state ($\delta = 1$). 
We compare the total mode involving all  ions (Sr-Eu-Mo-O) with three partial displacements: A-cation only mode (Sr-Eu), A-O mode (Sr-Eu-O) that includes the displacement of Sr, Eu and O atoms, and the B-O mode (Mo-O) whereby the Sr and Eu displacements are neglected.
As expected, the largest energy gain is obtained
when all atoms are included (Sr-Eu-Mo-O, \autoref{fig1}b). 
Since Mo cations remain largely centered in the octahedra, the results for the Mo-O mode show that oxygen displacements are essential to the structure stability; indeed, the polar Sr-Eu displacements alone lead to an increase in energy in the absence of O displacements (Sr-Eu, \autoref{fig1}b).
The difference between the total mode and that which omits the Mo displacements (Sr-Eu-O, Fig. \autoref{fig1}b) 
demonstrates that the ferroelectric ordering is stabilized by cooperative and coupled displacements of the Sr, Eu and O ions. We note that among all the O atoms the main contribution is due to apical anions found in the $A$O monoxide planes  (\autoref{fig1}a). 
The stability of these polar displacements depends on the delicate balance between electrostatic interactions of the local dipole moments they generate 
and the electronic screening effects in the crystal. 
We now consider the electronic properties of SEMO to address the role of screening.

\begin{figure}
\includegraphics[width=.95\columnwidth,angle=-0]{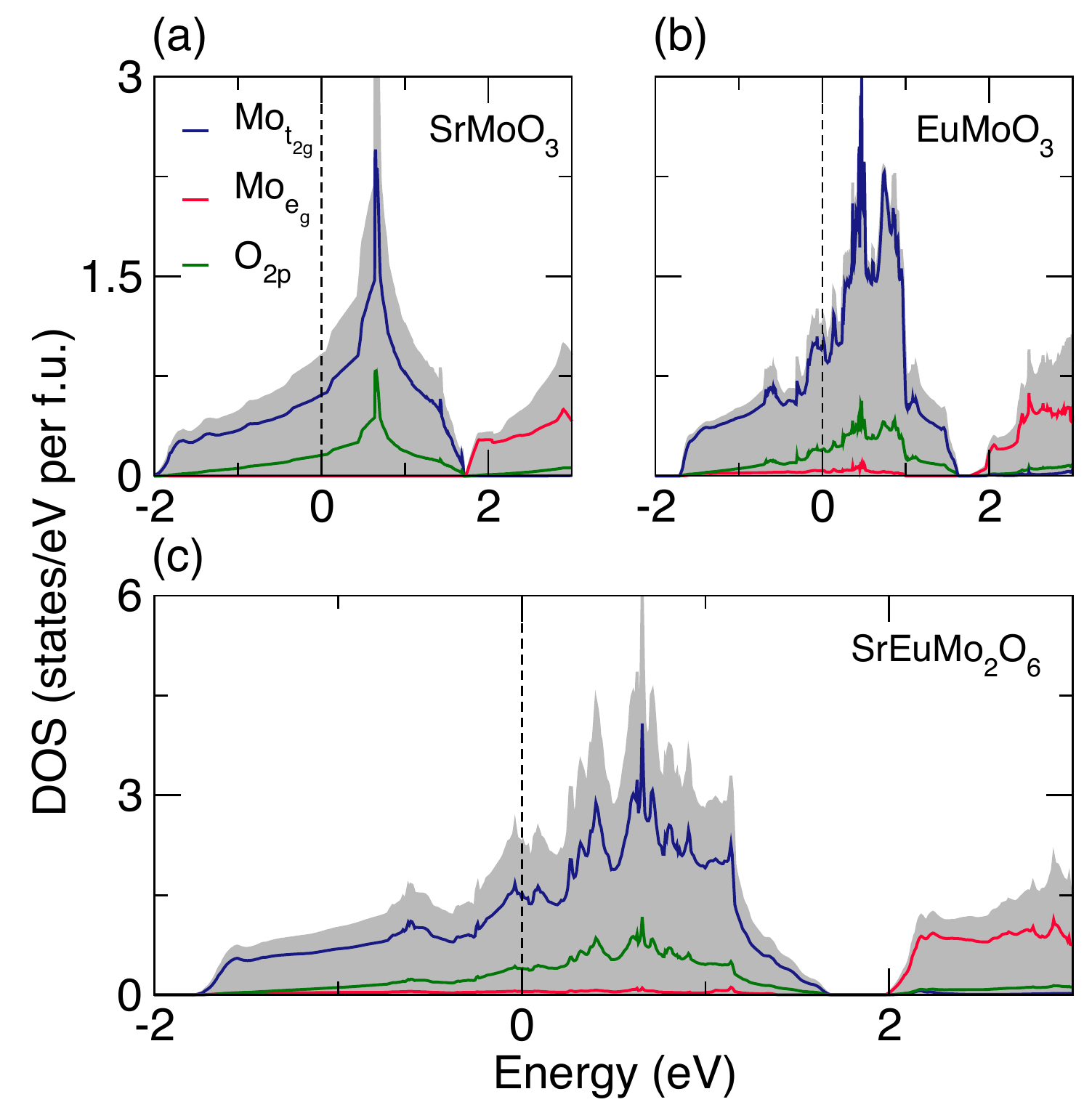}\vspace{-8pt}
\caption{(Color online) Atom- and orbital-resolved  resolved density of state for SMO (a), EMO (b) and SEMO (c) at the DFT-PBE level. The energy zero is set to the Fermi level.} 
\label{fig2}
\end{figure}

In \autoref{fig2} we compare the density of states (DOS) calculated within DFT-PBE 
for bulk SMO and EMO to that of SEMO.
The DOS of SMO and EMO are rather similar \cite{fstates}: Both bulk molybdates are found to be metallic with the extended Mo $4d$ states dispersing from approximately -2\,eV to 1.8\,eV above the Fermi level ($E_F$).
As shown in panel (c), SEMO is also a metal with sizable spectral weight at $E_F$. 
The low-energy contribution to the spectral density is dominated by the bands arising from the Mo $4d^2$ electrons, which are weakly entangled with the oxygen $2p$ states mainly  distributed from -8 eV to -4 eV below $E_F$ (not shown). 
The $4d$ bands in SEMO have an overall width of 3.4\,eV, which is reduced with respect to SMO and EMO owing to the increased tilting of the MoO$_6$ octahedra. The narrowing of the $t_{2g}$ manifold also separates it from the higher-lying $e_g$ bands.

\begin{figure}
\centering
\includegraphics[width=0.98\columnwidth]{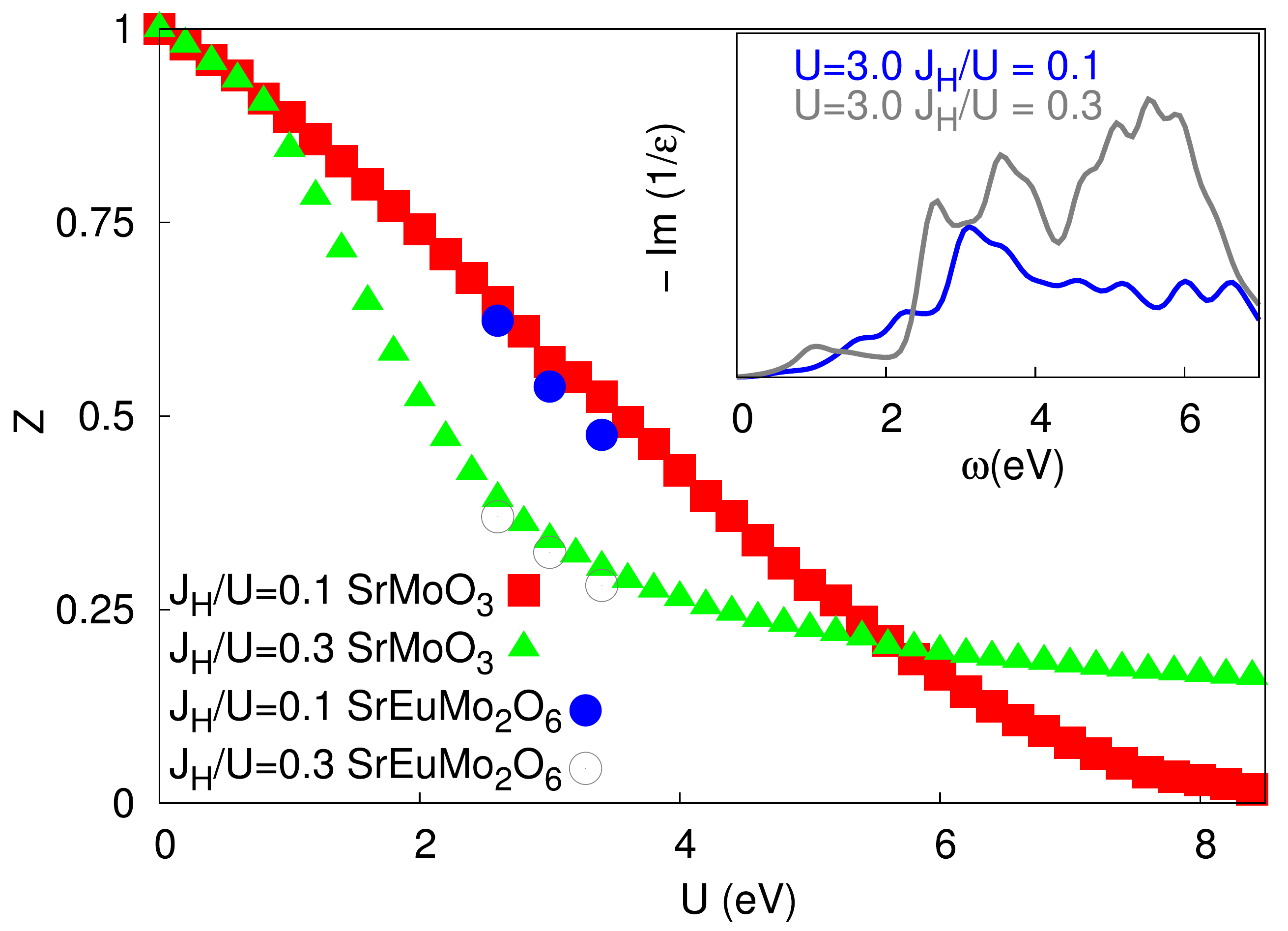}\vspace{-6pt}
\caption{(Color online) Quasiparticle weights ($Z$) calculated within DFT+DMFT for SMO, EMO and SEMO at different values of the ratio $J_H/U$. Inset: Imaginary part of the inverse dielectric function, $1/{\epsilon}$, for SEMO at different $J_H/U$ values. 
}
\label{fig3}
\end{figure}

Next we evaluate the screening behavior of the low-energy conducting electrons (\autoref{fig3}), investigating first electronic correlations in bulk SMO and then comparing its behavior with the SEMO superlattice. 
(Data for EMO is not shown because of the similarities in the results).
In order to include the on-site Coulomb interaction parameterized by the Hubbard $U$ and the Hund's coupling $J_H$ in our DMFT calculations (we use a Kanamori parametrization), we compute maximally-localized Wannier orbitals \cite{wannier90} for the $4d$ Mo states over the energy range spanned by the $t_{2g}$ orbitals across $E_F$ to construct the non-interacting part of our Hamiltonian.
In the DFT+DMFT  scheme \cite{DMFT}, which treats the lattice problem as an impurity embedded in a self-consistent bath, we employ Exact Diagonalization (ED) \cite{Caffarel,CaponeED} as the impurity solver using an Arnoldi algorithm \cite{ARPACK} to perform the diagonalization.

To characterize the degree of correlation of the system we study the quasiparticle weight $Z$, which is 1 for a non-interacting metal and decreases as a function of the interaction strength. A vanishing $Z$ signals a Mott insulating phase. The constrained random-phase approximation (cRPA) for bulk SMO gives $U = 3.0$\,eV and $J_H = 0.3$\,eV \cite{cRPA}.
In \autoref{fig3} we plot $Z$ as a function of $U$  for two values of the ratio $J_H/U$ ($J_H/U=0.1$, which corresponds to the cRPA values, and 0.3). For bulk SMO with cubic symmetry the $t_{2g}$ orbitals are perfectly degenerate and occupied by 2/3 electrons per orbital and the states share the same value of $Z$. For the cRPA value of $J_H/U=0.1$, we find $Z\simeq0.6$, consistent with experiment and previous theoretical estimates \cite{Wadati}.

The evolution of $Z$ follows the behavior discussed in Ref.~\onlinecite{MediciJanus}, where a rapid decrease for small $U$ is followed by a flattening of the $Z(U)$ curve, and ultimately to a Mott transition, which occurs at a relatively large value of $U$ ($\sim$8\,eV for $J_H/U = 0.1$). This behavior is more pronounced with increasing  $J_H/U$ with a faster initial decrease and a much slower decrease to an enhanced critical strength. For the cRPA estimates, we find that SMO is indeed on the brink of a `Hund's correlated' phase and far from the Mott state, which would require an unphysical enhancement of the ratio between $U$ and the bare kinetic energy.

We now turn to SEMO, where the loss of inversion symmetry alters the $t_{2g}$ manifold symmetry through a small crystal-field splitting and a change in the local hybridization, leading to small differences between the $Z$ values for the different orbitals. For the sake of simplicity, we plot the orbital-averaged $Z$, and note that the reduced symmetry introduces minor modulations of the orbital occupancies without changing the physical picture. Here we observe a small reduction of the quasiparticle weight $Z$ and consequently a slight increase of the degree of correlation (see the data plotted as circles in \autoref{fig3}), which pushes SEMO towards a correlated regime driven by the Hund's coupling.
Therefore the polar metallic state of SEMO is characterized by the simultaneous presence of polar distortions and a poor metallic behavior with intermediate electronic correlation strength.
Indeed, the correlated nature of the metallic state in bulk SMO is evidenced experimentally by a low quasiparticle coherence scale around $T^*= 140$\,K \cite{Nagai}, which is of the same order as Sr$_2$RuO$_4$\cite{Wang2} above which highly incoherent conduction characterizes the system. Based on our DMFT calculations, we expect similar behavior for SEMO. Above $T^*$ the incoherent conduction electrons would contribute even less to the screening of local dipole moments formed by the polar lattice mode, further favoring the coexistence of the broken inversion symmetric state and metallicity. 

To further justify this argument, we also plot the contribution of the Mo~$4d$ electrons to the frequency-dependent imaginary part of the inverse dielectric function ($1/{\epsilon}$) for SEMO, which directly captures the effects of screening (\autoref{fig3}, inset) \cite{Nota}.
Although the full screening of the Coulomb interactions in a crystal is due to contributions from all orbitals, \emph{e.g.}, including those of the ligands, we find that in correlated SEMO the contribution from the conduction orbitals is strongly reduced owing to electron-electron correlations which are strongly enhanced by increasing the Hund's coupling.
On average $\Im(1/\epsilon)$  is larger for $J_H = 0.3$\,eV in the more correlated regime owing to the reduced $Z$ derived from Hund's assisted correlations. 
Both features give a reduction to dielectric screening and in part alleviates the incompatibility between metal and ferroelectric-like orders. 
%

The present study, together with previous indications of strong correlations in LiOsO$_3$ \cite{GGMC} and other NCSM, suggests that materials with $d^2$ or $d^4$ configurations are promising candidates from which to realize new polar metals. Provided ferroelectric-like displacements can be realized through a lattice instability, poor screening of the electric dipoles will be provided in the correlated metallic state  when the Hund's coupling is sufficiently large. An interesting example could be given by LiReO$_3$, which displays the same polar crystal class as LiOsO$_3$ ($R3c$) \cite{Cava} owing to the off-centering of Li ions, while it shares the same nominal $d^{2}$ electronic configuration of SEMO. 
Finally we also remark that the proposed  ultra-short NCSM ruthenate (SrRuO$_3$)$_1$/(CaRuO$_3$)$_1$ superlattices  have a $d^4$ Ru$^{4+}$  configuration \cite{PuggioniRondinelli}. 
The ruthenate electronic structure, analogously to the $d^2$  of SEMO, is ideal for  Hund's metal physics, as has been discussed for the bulk end-members  \cite{MravljeIncoherent,MravljeRu}. Therefore the electronic stability of the NCSM state in this superlattice may be supported by the same mechanism described herein, \emph{i.e.}, poor metallicity assisted by Hund's coupling.

In summary, we propose SrEuMo$_2$O$_6$ as polar Hund's metal. Our electronic structure calculations reveal that the polar structure is driven by 
asymmetric displacements of the ions in the SrO and EuO planes, which occur with  ferromagnetic ordering of spins on the Mo ions.
The ferroelectric-like state coexists with a correlated metallic phase which relies on the Hund's coupling and it is robust with respect to Mott localization. 
The poorly coherent quasiparticles are ineffective at completely screening the 
ordered local dipole moments. 
These results strengthen the link between correlated metallic states and the propensity of 
a metallic material to adopt polar structure, facilitating the selection or design of new 
correlated polar metals for enhanced magnetoelectronic responses or customized 
antisymmetric exchange interactions, which support exotic magnetic textures (helical or skyrmionic structures). 
Although our current understandinding of polar metals reveals that most of them 
are in a correlated electronic regime with bad metallic behavior, we do not eliminate 
the possibility that non-correlated polar metals may exist; rather, strong 
electronic correlations appears to be a favorable ingredient to stabilizing noncentrosymmetric metallic phases, but it may not be a prerequisite in scenarios where  
the inversion lifting displacements are indeed  decoupled from the low-energy electronic structure.
%

G.G.\ and M.C.\ acknowledge financial support from the European Research Council
under FP7/ERC Starting Independent Research Grant ``SUPERBAD" (Grant No.\ 240524). 
D.P.\ and J.M.R.\ acknowledge the Army Research Office for financial support (Grant No.\ W911NF-15-1-0017), and the 
HPCMP of the DOD and  
XSEDE supported by NSF (Grant No.\ OCI-1053575) for providing computational resources. 




\begin{thebibliography}{99}
\bibitem{note1} A kinetic cutoff energy of 550 eV is used to expand the wavefunctions and a $\Gamma$ centered $10\times8\times10$ $k$-point mesh combined with the tetrahedron method is used for Brillouin zone integrations. Atomic relaxations are continued until the changes in the total energy are less than $10^{-8}$\,eV.
\bibitem{relaxed_structure} Using the lattice vectors: $(a,b,c)$=$(5.621, 7.983, 5.621)$\, \AA, the optimized Wyckoff positions are: 
Sr: $2a, z=0.497$; 
Eu: $2b, z=0.508$; 
Mo: $4d, y=0.75, z=0$; 
O(1): $8e, x=0.747, y=0.778, z=0.246$; 
O(2): $2b, z=0.059$; and 
O(3): $2a, z=-0.246$.
\bibitem{fstates}  The $f$ electrons of Eu are kept frozen in our DFT calculations being localized at 2 eV below the Fermi level in the calculations of EuMoO$_3$ \cite{Kozuka}.
\bibitem{Nota} The imaginary part of the dielectric function is obtained from the DMFT  optical conductivity ${\sigma}(\omega)$ as ${{\epsilon}_2}(\omega)$=4$\pi$${\sigma}(\omega)$/$\omega$, and the real part  may be obtained via the Kramers-Kronig relation to provide the full complex dielectric function.
\end{thebibliography}
\end{document}